\documentclass[12pt]{spieman}  
 
\usepackage{amsmath,amsfonts,amssymb}
\usepackage{graphicx}
\usepackage{setspace}
\usepackage{tocloft}
\usepackage[colorlinks=true, allcolors=blue]{hyperref}
\usepackage{threeparttable,subcaption}
\usepackage{comment}
\usepackage{here}
\usepackage{siunitx}
\usepackage{multirow}
\usepackage{color}

\title{Ground test results of the micro-vibration interference for the x-ray microcalorimeter onboard XRISM}
\author[a,*]{Takashi Hasebe}
\author[b]{Ryuta Imamura}
\author[c]{Masahiro Tsujimoto}
\author[b]{Hisamitsu Awaki}
\author[d]{Meng P. Chiao}
\author[e]{Ryuichi Fujimoto}
\author[d]{Leslie S. Hartz}
\author[d]{Caroline A. Kilbourne}
\author[d]{Gary A. Sneiderman}
\author[c]{Yoh Takei}
\author[f]{Susumu Yasuda}

\author[]{the XRISM \textit{Resolve} team}

\affil[a]{Kavli Institute for the Physics and Mathematics of the Universe (WPI), The University of Tokyo, Chiba, Japan}
\affil[b]{Graduate School of Science and Engineering, Ehime University, Ehime, Japan}
\affil[c]{Institute of Space and Astronautical Science, Japan Aerospace Exploration Agency (JAXA), Kanagawa, Japan}
\affil[d]{National Aeronautics and Space Administration (NASA), Goddard Space Flight Center, MD, USA}
\affil[e]{Faculty of Mathematics and Physics, Kanazawa University, Ishikawa, Japan}
\affil[f]{Japan Aerospace Exploration Agency (JAXA), Ibaraki, Japan}

\cftpagenumbersoff{figure}
\cftpagenumbersoff{table} 
\begin{document} 
\maketitle

\begin{abstract}
 \textit{Resolve} is a payload hosting an X-ray microcalorimeter detector operated at
 50~mK in the X-Ray Imaging and Spectroscopy Mission (XRISM). It is currently under
 development as part of an international collaboration and is planned to be launched in
 2023. A primary technical concern is the micro-vibration interference in the sensitive
 microcalorimeter detector caused by the spacecraft bus components. We conducted a
 series of verification tests in 2021--2022 on the ground, the results of which are
 reported here. We defined the micro-vibration interface between the spacecraft and the
 \textit{Resolve} instrument. In the instrument-level test, the flight-model hardware
 was tested against the interface level by injecting it with micro-vibrations and
 evaluating the instrument response using the 50 mK stage temperature stability, ADR
 magnet current consumption rate, and detector noise spectra. We found strong responses
 when injecting micro-vibration at $\sim$200, 380, and 610 Hz. In the former two cases,
 the beat between the injected frequency and cryocooler frequency harmonics were
 observed in the detector noise spectra. In the spacecraft-level test, the acceleration
 and instrument responses were measured with and without suspension of the entire
 spacecraft. The reaction wheels (RWs) and inertial reference units (IRUs), two major
 sources of micro-vibration among the bus components, were operated. In conclusion, the
 observed responses of \textit{Resolve} are within the acceptable levels in the nominal
 operational range of the RWs and IRUs. There is no evidence that the resultant energy
 resolution degradation is beyond the current allocation of noise budget.
\end{abstract}


{\noindent \footnotesize\textbf{*}Takashi Hasebe,  \linkable{takashi.hasebe@ipmu.jp} }

\begin{spacing}{1}   

\section{Introduction}\label{s1}
The X-Ray Imaging and Spectroscopy Mission (XRISM) is an x-ray observatory planned to be
launched in 2023 \cite{Tashiro2020a}. One of the scientific payloads is
\textit{Resolve}\cite{ishisaki2022}, which hosts an x-ray microcalorimeter for high
resolution spectroscopy. It was developed under the collaboration of JAXA, NASA, ESA,
and universities and institutes under these agencies. \textit{Resolve} is based on the
soft x-ray spectrometer (SXS)\cite{Kelley2016,Mitsuda2014} onboard
ASTRO-H\cite{takahashiJATIS} that was launched in 2016 February. The SXS achieved a
stable in-orbit spectroscopic performance with an energy resolution of 5 eV FWHM at 5.9
keV\cite{porter18}. The success was suddenly terminated by the loss of the spacecraft
attitude control in a month. \textit{Resolve} was made with almost the same design as
the SXS to recover its scientific programs.

Micro-vibration interference is a primary concern in the design of instruments hosting a
low-temperature detector, which is sensitive to all forms of energy inputs dissipating
into heat. In particular, microcalorimeters or bolometers that use conventional
high-impedance semiconductor detectors coupled with mechanical cryocoolers in the
cooling chain are susceptible to this interference.  Examples include the
Planck\cite{Tauber2010} HFI\cite{lamarre10} using neutron-transmuted doped Ge of
$\sim$10~M$\Omega$ operated at 100 mK, and the ASTRO-H SXS and XRISM \textit{Resolve}
using ion-implanted Si of $\sim$30~M$\Omega$ at 50 mK. In fact, strong pickup lines were
observed in the detector noise spectra both in the HFI and SXS at the driving frequency
of the cryocoolers (15–52 Hz) and their harmonics.

In SXS, the cryocooler micro-vibration degraded detector performance to an unacceptable
level. The temperature fluctuation at the 50 mK stage was $\sim$40~$\mu$K rms, whereas
2.5~$\mu$K was budgeted. As a countermeasure, a significant change was made in the
hardware at a very late stage of development. The vibration isolation system (VIS) was
newly designed and installed between the cryocooler compressors and cryostat a few
months before the final integration into the spacecraft\cite{takei18}. This
countermeasure was successful and reduced the fluctuation to 0.5 $\mu$K rms. For
\textit{Resolve}, the VIS between the cryocoolers and the cryostat is included from the
beginning. It has an improved design that includes a launch-lock mechanism and has
better dumping performance\cite{Ezoe2020}. The micro-vibration interference by the
cryocoolers to the detector is discussed in a separate paper\cite{imamura2022,imamura_jltp}.

In addition to the cryocoolers in the instrument, the spacecraft also hosts potential
micro-vibration sources, such as reaction wheels (RW) and inertial reference units (IRU)
in the attitude control system. In the case of SXS, these components exhibited some
hints of micro-vibration interference during the spacecraft test. However, the testing
time on the ground and the life in orbit were insufficient to assess their impact and
isolate the problem. Spacecraft interference poses a challenge in the
integration program as it can be identified only after the instrument is integrated into
the spacecraft when limited resources remain for major changes. For \textit{Resolve}, a
micro-vibration control plan was formulated and a series of tests were performed to
control the risk. The purpose of this article is to describe the results obtained in the
series of ground tests from 2021 to 2022 at JAXA's Tsukuba Space Center using flight
model hardware. We aim to provide a case study of the micro-vibration control of a
space cryogenic mission, which will be followed by many others.

The remainder of this paper is organized as follows. In \S~\ref{s2}, we give a brief
description of the \textit{Resolve} instrument and the spacecraft bus system, and the
micro-vibration interface between them. 
The sensitivity of \textit{Resolve} to micro-vibration was tested before integration onto the spacecraft through the injection of micro-vibrations at simulated interfaces to the spacecraft (\S~\ref{s3}).
The actual level of micro-vibration at the interface by the
spacecraft was measured and an end-to-end assessment from the bus components to the
detector was made in the spacecraft-level test (\S~\ref{s4}). We summarize the result in
\S~\ref{s5}.

\section{Hardware and Interface}\label{s2}
\subsection{Spacecraft}\label{s2-1}
\begin{figure}[htbp]
 \centering
 \includegraphics[scale=0.6]{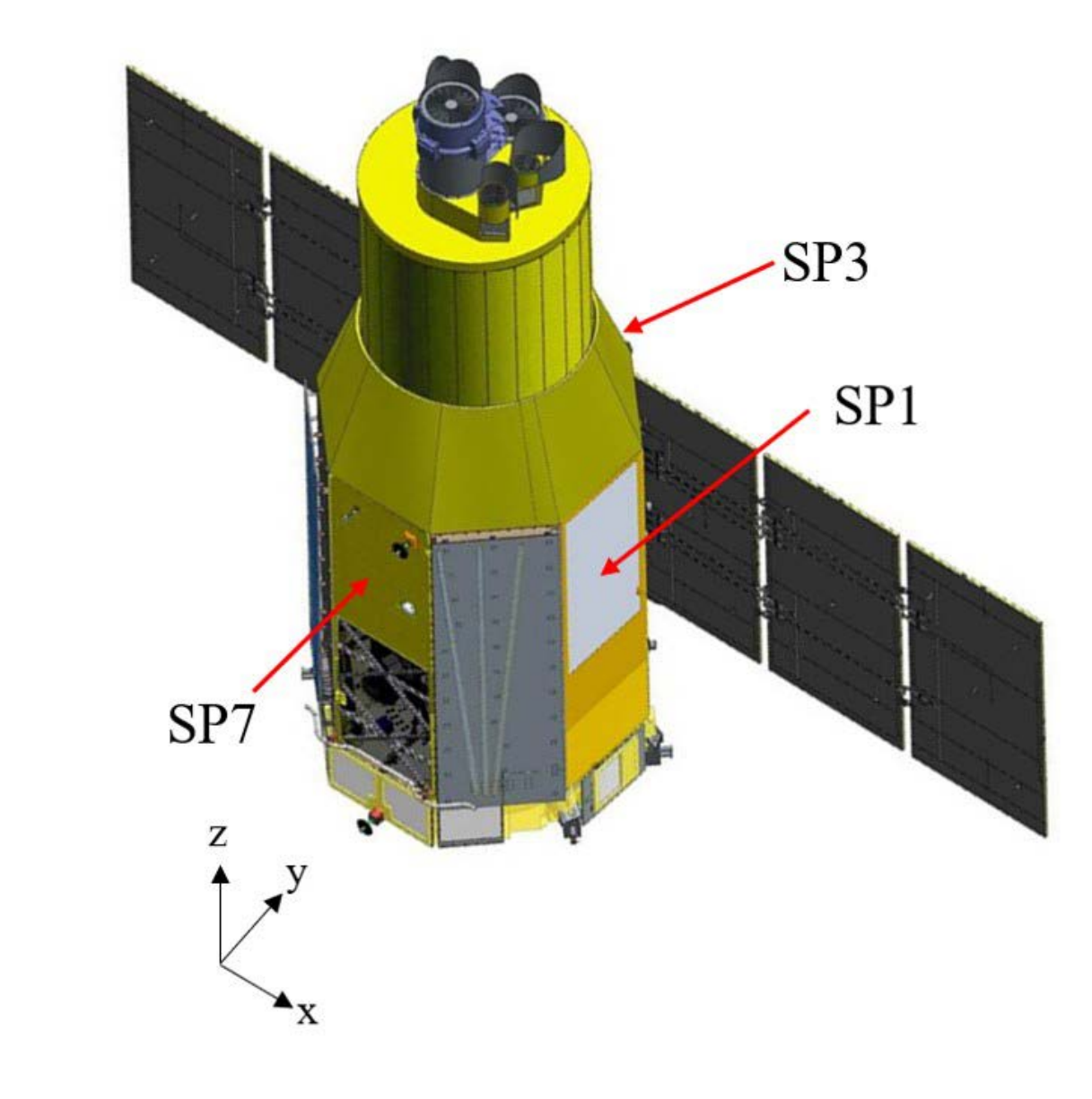}
 \includegraphics[scale=0.4]{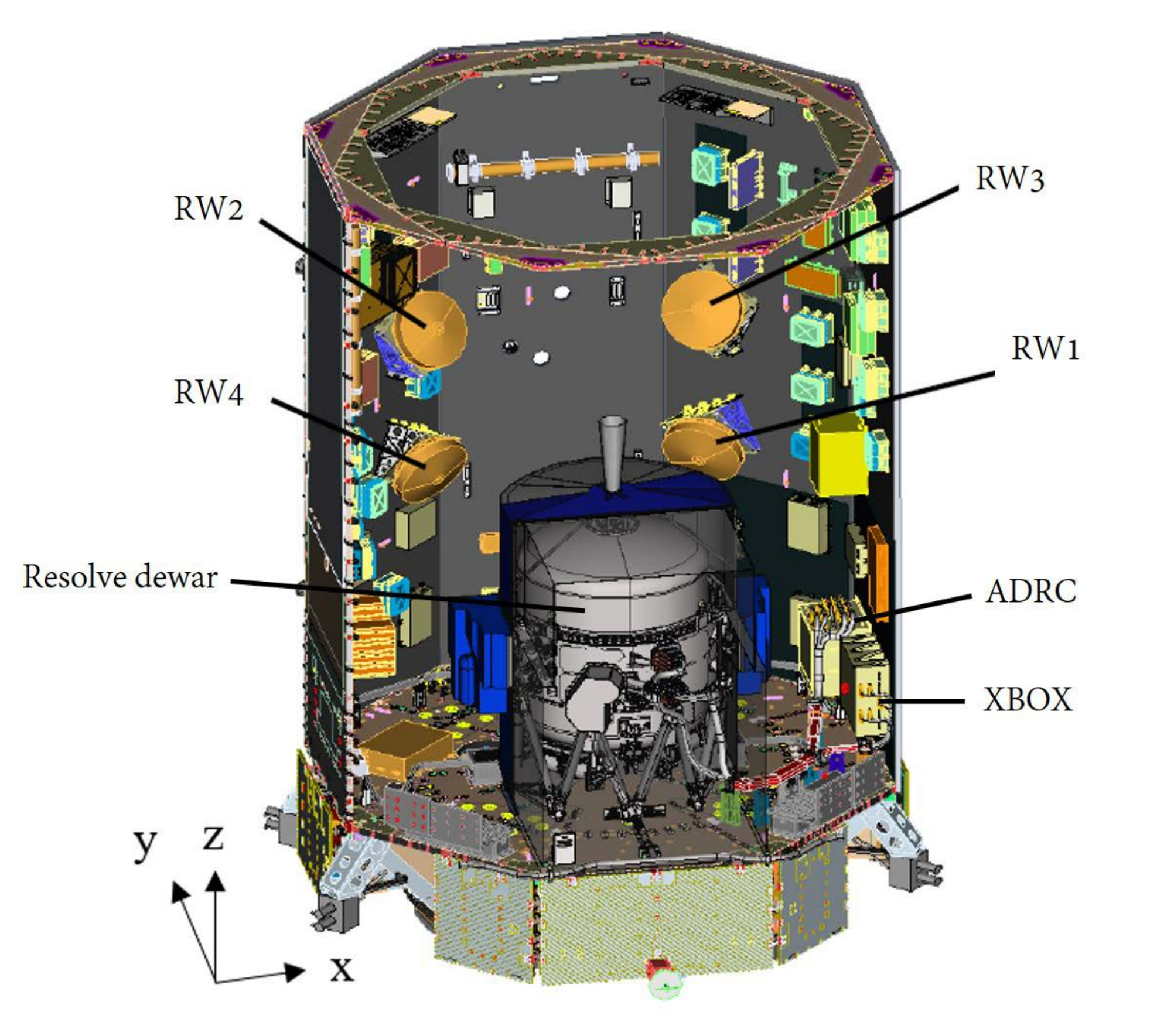}
 \caption{Schematic view of the spacecraft (left) and cross-sectional view inside, in
 which the lower and middle plates and struts are removed (right). The drawing is
 provided by NEC.}
 \label{fig:xrism}
\end{figure}

The spacecraft design of XRISM is based on ASTRO-H's. NEC is the prime contractor for
both. The spacecraft will be put into a near-earth orbit at an altitude of 575 $\pm$
15~km and an inclination angle of 31~degrees. A schematic view is shown in
Figure~\ref{fig:xrism} left. It has a size of $7.9 \times 9.2 \times 3.1$~m$^{3}$ and a
weight of $2.3 \times 10^{3}$~kg\cite{Tashiro2020a}. The main body is made of eight side
panels (SP1--8) of $990 \times 3100$~mm$^{2}$ and the base, lower, middle, and top
plates perpendicular to the side panels. X-ray telescopes and the star trackers are
placed on the top plate, while \textit{Resolve} dewar is on the base plate. The solar
panel is attached in parallel with the SP3 toward the Sun, while the dewar is exposed to the deep
space through the open space on the opposite side (SP7) for radiative cooling.

The spacecraft has four RWs (RW1--4) to maintain the pointing during observations and
three magnetic torquers to dump the accumulated angular momentum against the Earth's
magnetic field. The attitude is measured by four independent sensors: IRUs, the
geomagnetic aspect sensors, the digital sun sensors, and the star trackers. These
actuators, sensors, and their driving electronics are stored inside the side panels
(Figure~\ref{fig:xrism} right). The RWs and IRUs are the two major sources of
micro-vibration in the spacecraft system. The RWs are operated for the three-axis
control at a nominal rotation speed of 1.5--4.5~krpm (25--75~Hz) with the base at
3.0~krpm (50~Hz). Two (RW1 and 3) are located inside of the SP2 and the other two (RW2
and 4) in SP4. The IRUs consist of two sets (IRU-A and B) of two tuned dry gyros, and
three of them are used at a fixed rotation speed of 155~Hz. The IRUs are stored in the
spacecraft base panel to avoid thermal distortion of the side panels. 
These frequencies are right in the middle of the microcalorimeter bandpass of 12 to a few hundred Hz.

\subsection{Instrument}\label{s2-2}
The x-ray microcalorimeter detector has an array of 6$\times$6 pixels that are thermally
anchored to the 50~mK bath with a thermal time constant of a few ms\cite{porter18}. The
detector is placed inside the dewar, which consists of several layers of shields made of
Al for thermal insulation and vacuum\cite{fujimoto18,Yoshida2018}. Inside the dewar, the
superfluid He is stored in the He tank. Two-stage adiabatic demagnetization refrigerator
(ADR) is used to pump heat from the 50 mK stage to the He tank\cite{shirron18}. The
evaporated He is used to vapor-cool the shields inside the dewar. When the He runs out,
a third ADR is used to cool down the He tank to maintain the 50 mK control for an
extended lifetime\cite{sneiderman18,Kanao2017}. On the surface of the dewar, five
cryocoolers are attached for active cooling of the shields to reduce the thermal load
toward the coldest stage. Two Stirling coolers (STC)\cite{Sato2012} cool the 100 and
30~K stages, while one Joule-Thomson cooler (JTC) cool the 4~K stage. The other two STCs
are used for pre-cooling the JTC. The STC’s have a compressor containing two opposing pistons and are driven at 15 Hz, while the JTC has two compressors containing two
opposing pistons to generate low and high pressures and is driven at 52 Hz. The frequency
can be tuned by $\pm$1~Hz for STC and $\pm$2~Hz for JTC by commands to avoid possible
resonances.  Each pixel of the detector is read out independently at a sampling rate of
12.5~kHz. The readout electronics called the XBOX\cite{Kelley2016} supplies the detector
bias, and samples, shapes, amplifies, and digitizes the detector signal. The ADR
controller (ADRC\cite{Kelley2016}) controls the ADR magnet current and the heat switches
as well as the thermal control of the cold stages inside the dewar. These two
room-temperature electronics have wires reaching deep inside the dewar, thus are
considered to be possible routes of micro-vibration into the detector. Both are placed
inside of the SP1 (Figure~\ref{fig:xrism} right).

\subsection{Interface}\label{s2-3}
\begin{figure}[htbp]
 \centering
 \includegraphics[scale=0.65]{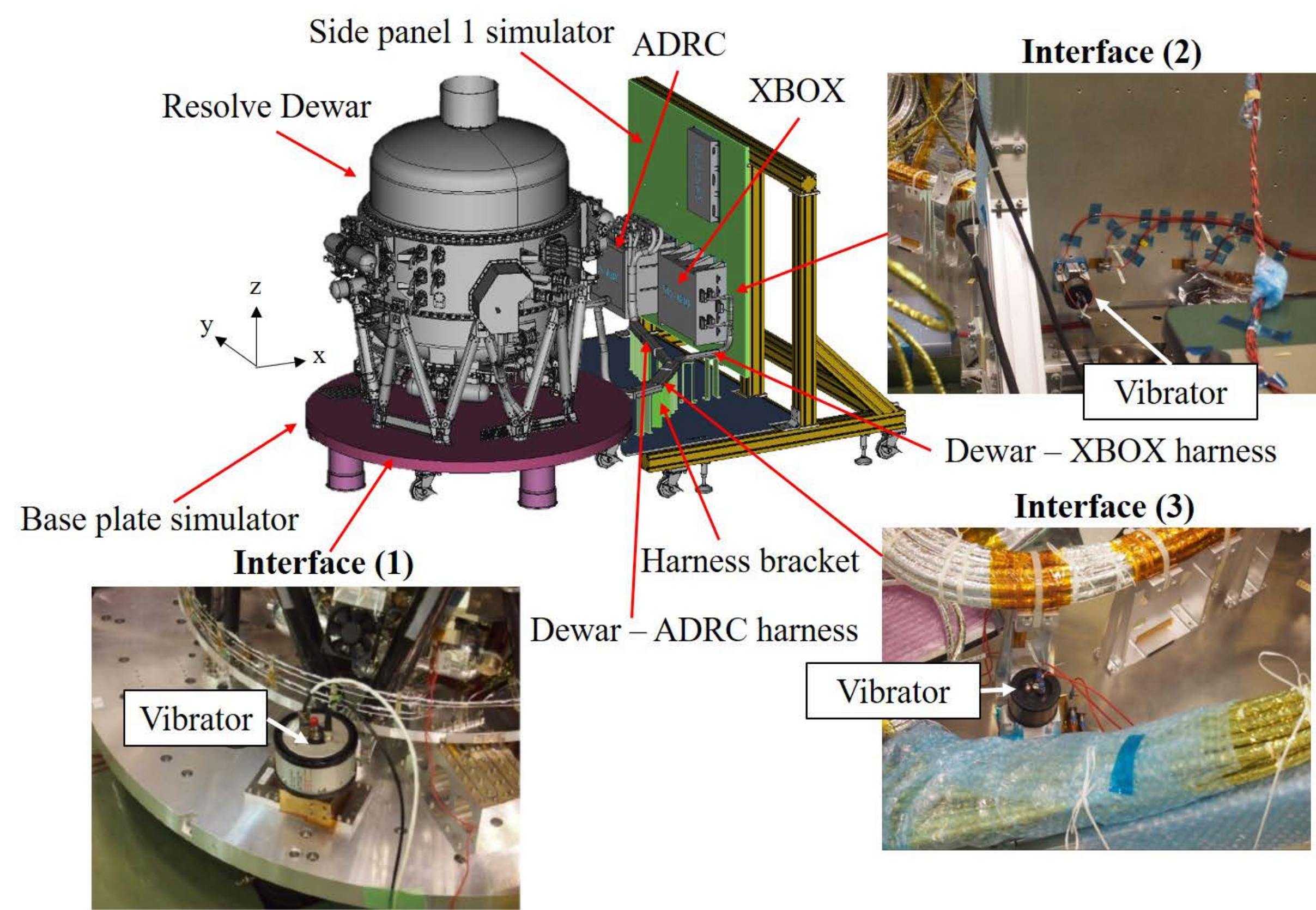}
 \caption{Schematic view of the spacecraft--\textit{Resolve} interfaces and the photos
 of vibrators used in the instrument-level test. The drawing is provided by NEC.}
 \label{fig:interface}
\end{figure}

We defined the micro-vibration interface between the spacecraft and the instrument at
three positions (Figure~\ref{fig:interface}): (1) The spacecraft base plate, on which
the dewar stands with its eight struts. 
They are the only points that the dewar is mechanically supported on the spacecraft.
(2) The SP1, at the foot of XBOX, where the
XBOX and ADRC are located. (3) The harness connecting the dewar and the XBOX and
ADRC. The harness is supported by the brackets on the base plate, which are defined as
an interface. 
For each interface, the maximum allowed vibration input as a function of frequency was estimated based on the allocation for the micro-vibration interference in the \textit{Resolve} noise budget. This limit was then imposed as a requirement on the bus components.
The levels are defined in acceleration (m~s$^{-2}$) as a function of frequency in the $z$
axis for (1) and (3) and $x$ axis for (2) perpendicular to SP1. The interface levels and
frequency ranges (Figure~\ref{f04}) were given based on the SXS results for (1) and on a
component-level test for (2). The stricter of the two was applied for (3).

\section{Instrument-level tests}\label{s3}
\subsection{Measurements}\label{s3-1}
\begin{figure}[!hbtp]
 \centering
 \includegraphics[scale=0.7]{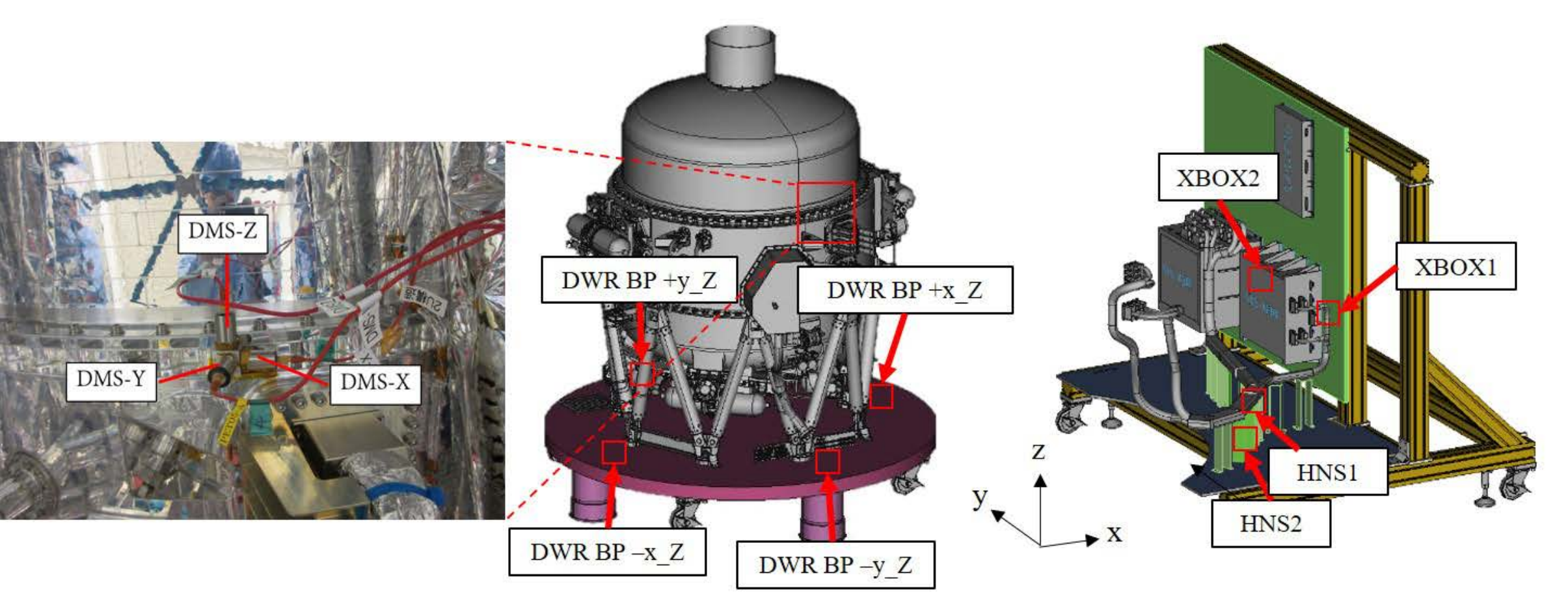}
 \caption{Positions of the accelerometers and their identifications. The XBOX1 and XBOX2
 accelerometers are placed on the backside of the SP1 simulator. The drawing is provided
 by NEC.}
 \label{fig:coordinate}
\end{figure}

We conducted the instrument-level test in 2021 December 4--11. The \textit{Resolve}
instrument was operated in the nominal configuration. The drive
frequencies of the STC and JTC were 14.452 and 53.478~Hz, respectively. We fabricated
the spacecraft interface simulators for the base plate, the harness support brackets,
and the SP1, where the interfaces between the spacecraft and the instrument are
defined. Note that there are some differences; the base plate and the SP1 simulators are
made of Al, while the flight equivalent units are of Al honeycomb with the CFRP skin. At
each interface,
we attached a vibrator with a force sensor for injecting micro-vibration
(Figure~\ref{fig:interface}) and piezoelectric accelerometers for monitoring levels
(Figure~\ref{fig:coordinate}).
The accelerometers were read out continuously at a
1.6~kHz sampling rate with a low pass filter cut off at $\sim$800~Hz. On the base plate
simulator, a Wilcoxson F4 vibrator was placed at a position displaced from the center in
the $+y$ direction and four accelerometers 90 degrees apart 
(DWR BP $\pm x\_Z$ or $\pm y\_Z$) were placed along the $z$ axis.
On the back side of the SP1 simulator, an F3 vibrator and two sets of three-axis accelerometers were attached.
Close to the bracket, another F3 vibrator and two other sets of accelerometers for the three axes were placed.

\begin{figure}[!hbtp]
 \centering
 \subcaptionbox{Interface (1) Base plate\label{fig:dwr_sweep}}
 {
 \includegraphics[scale=0.35]{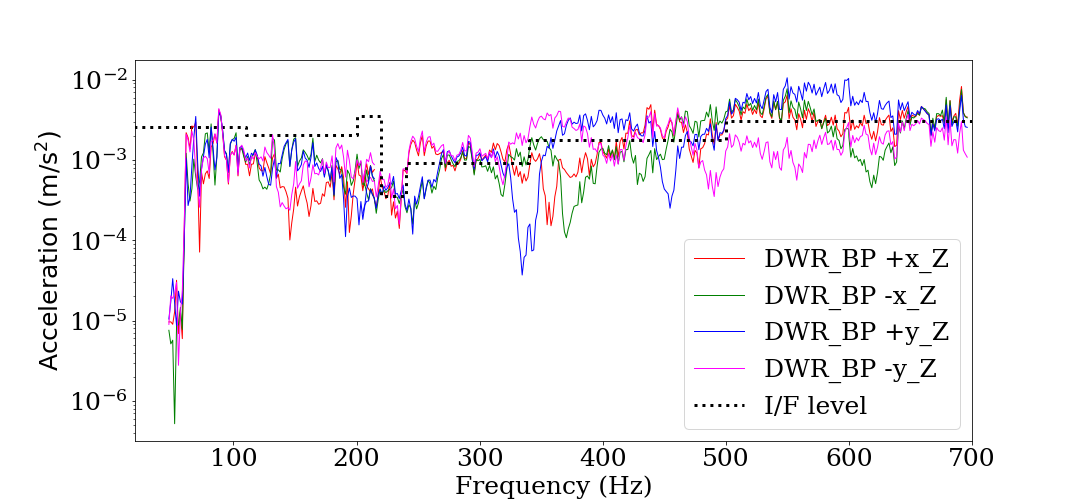}
 }
 \subcaptionbox{Interface (2) Side panel 1\label{xbox_sweep}}
 {
 \includegraphics[scale=0.35]{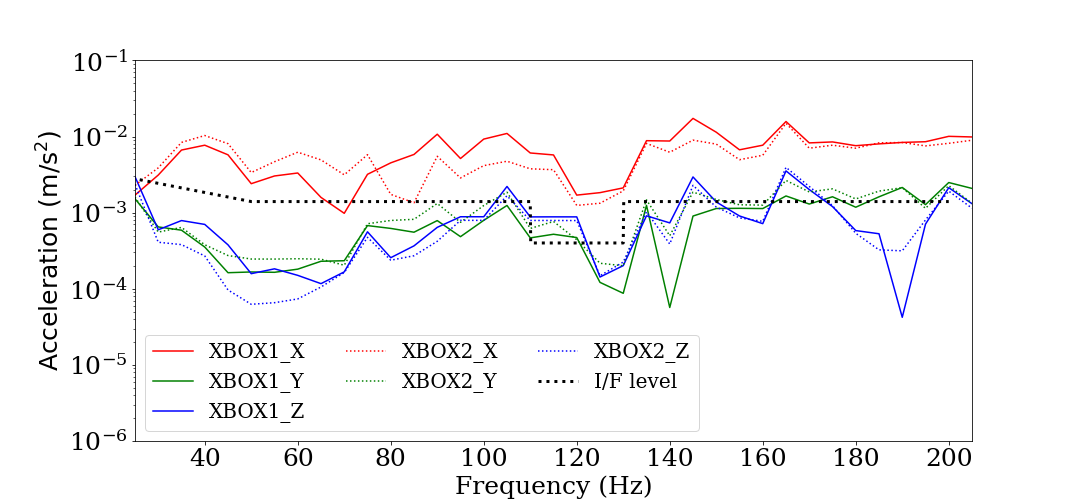}
 }
 \subcaptionbox{Interface (3) Harness support bracket\label{hns_sweep}}
 {
 \includegraphics[scale=0.35]{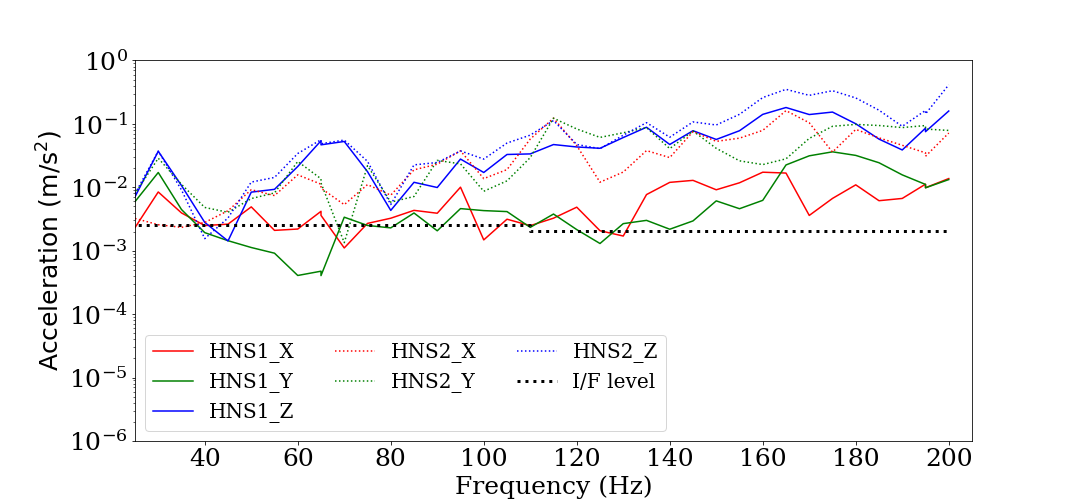}
 }
 \caption{Interface levels and the injected levels during the frequency sweep.}
 \label{f04}
\end{figure}

For each interface, we conducted three runs: 
(i) force calibration, (ii) frequency sweep, and (iii) dwell. For (i),
both the force and accelerometer sensors were read by giving a fixed force over the
interface frequency range, from which the force necessary to achieve the interface level
of the acceleration was derived as a function of frequency. For (ii), we swept over the
frequency at the derived force (Figure~\ref{f04}) and evaluated the excess heat load to
the detector and its stability using the ADR magnet current consumption rate
($dI_{\mathrm{ADR}}/dt$) and the 50 mK temperature fluctuation ($T_{\mathrm{rms}}$)
sampled at 1~Hz. Two neutron transmutation doped Germanium resistance
thermometers are used for the 50~mK stage; one is for the PID control of the magnet
current and the other for monitor. These thermometers are known to exhibit glitches in
reading, thus we judged the positive response only when at least two of the three
measurements (two $T_{\mathrm{rms}}$ and one $dI_{\mathrm{ADR}}/dt$ values) agreed. We
tried injecting the interface level or beyond as accurately as possible, but ended up
injecting lower levels in some frequencies, in particular in the low-frequency range,
due to the power limitation of the vibrators. For such frequencies, we made additional
dwell measurements for linearity check. For (iii), we dwell at selected frequencies
based on the sweep results for 5~min to collect the detector noise spectra. We also
changed the amplitudes of the micro-vibration injection to examine the linearity of the
response.

\subsection{Results}\label{s3-2}
\begin{figure}[htbp]
 \centering \includegraphics[width=0.9\textwidth]{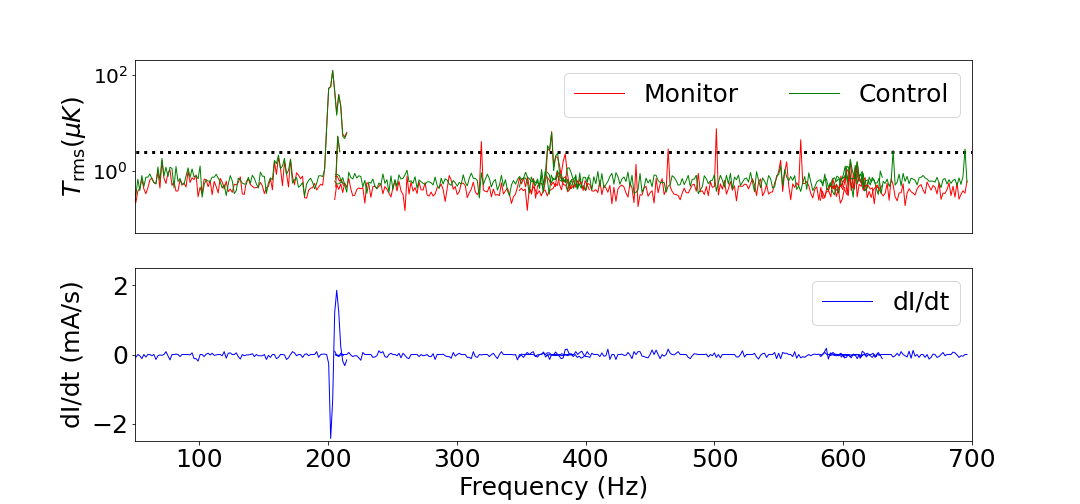} 
 \caption{Stability of the 50 mK stage thermometers ($T_{\rm{rms}}$; top) and the ADR
 magnet current consumption rate ($dI_{\mathrm{ADR}}/dt$; bottom) during the frequency sweep
 injection from the base plate. The horizontal dashed line in the top panel indicates
 the acceptable level of 2.5~$\mu$K rms. The result was obtained by injecting a level
 higher than that defined in figure~\ref{f04} (a) to investigate possible margins.}
 \label{fig:dwr_temp}
\end{figure}

We first examine the $T_{\mathrm{rms}}$ values against the acceptable level of
2.5~$\mu$K rms for the three interface injections. No deviation was found for the
interface 2 (SP1). The result of the interface 3 (harness support bracket) was
encompassed in that of the interface 1 (base plate). We thus focus on the result of the
base plate injection. Figure~\ref{fig:dwr_temp} shows the $T_{\mathrm{rms}}$ and
$dI_{\mathrm{ADR}}/dt$ responses as a function of the injection frequency. The positive
responses were found at $\sim$200, 380, and 610~Hz. The one at $\sim$200~Hz is far
beyond the acceptable level of $T_{\mathrm{rms}}$.

The former two frequencies are of particular interest as they match with the 4'th and
7'th harmonics of the JTC drive frequency, at which peculiar responses in the detector
noise spectra were observed during the JTC frequency tuning in different
experiments\cite{imamura2022,imamura_jltp}. We did not observe such responses in other JTC
harmonics. The peculiar response is the presence of the low-frequency beat between 4'th
harmonics of the JTC and 14 or 15'th harmonics of the STC for $\sim$200~Hz and between
7'th of the JTC and 24'th or 25'th of the STC for $\sim$380~Hz. We speculate that the
instrument has a resonance at these frequencies in cold stages, where the two harmonic
lines of the JTC and STC are mixed and amplified non-linearly to redistribute some power
into the beat frequency at $<$20~Hz. 
They were found to be dissipated somehow into heat at the cold stage, as the beat frequency noise was not observed when the detector bias was off to be insensitive to heat\cite{imamura2022,imamura_jltp}. The elevated levels of $T_{\rm{rms}}$ and $dI_{\mathrm{ADR}}/dt$ is considered to be related to these
resonances.

\begin{figure}[htbp]
 \centering
 \includegraphics[scale=0.35]{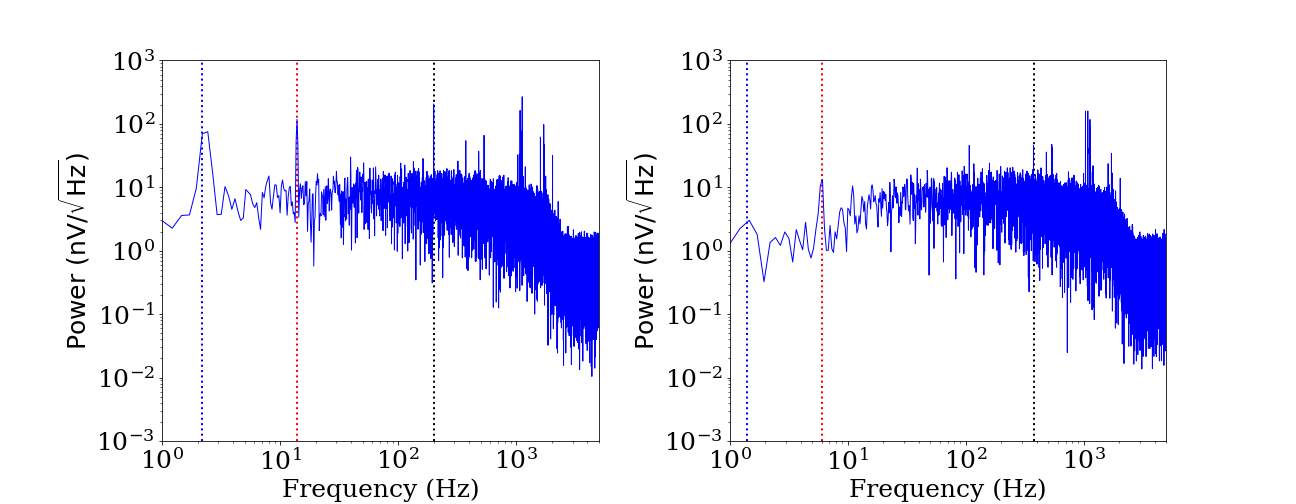}
 \caption{Noise spectrum of the pixel 0 of the microcalorimeter. (left) 200 Hz injection
 from the harness support bracket and (right) 380.4 Hz injection from the base
 plate. The black vertical line indicates the frequency of injection, while the blue and
 red lines are the beat frequencies among the injected frequency and STC or JTC harmonic
 frequencies.}
 \label{fig:low_noise}
\end{figure}

The low-frequency lines such as this degrade the detector performance
significantly. Before this test took place, we tuned the JTC and STC frequencies so that
they do not cause a low-frequency beat. However, during the micro-vibration test, a
third line was injected at these resonant frequencies, which was mixed with the existing
JTC and STC harmonic lines. We obtained the detector noise spectra during the
micro-vibration injection of these frequencies (Figure~\ref{fig:low_noise}). The
low-frequency lines are evident, which are attributable to the beats among the injected
frequency and the existing JTC and STC harmonics. In the 200~Hz injection (left), the
line at 2.2~Hz is the beat between the injected frequency and the 14'th STC harmonic,
and the one at 13.9~Hz is between the injected frequency and the 4'th JTC harmonic. In
the 380.4~Hz injection (right), the line at 1.4~Hz is the beat between the 7'th JTC and
24'th STC harmonic and the one at 6.1~Hz is between the injected frequency and the 7'th
JTC harmonic.

We are now concerned that the spacecraft system may bring the third line at the
frequencies of the enhanced sensitivity, particularly at $\sim$200~Hz with an
unacceptable level of response. The drive frequencies of the RWs (25--75~Hz) and the IRUs
(155~Hz) are lower than this, but their non-diagonal redistribution component into
higher frequencies may hit the sensitive frequencies. In the RW unit test, such
a non-diagonal component was indeed found. The coupling between the eigen frequency of the
RW at rest ($\omega_0$) and the rotating frequency ($\Omega$) causes precession
frequencies at
\begin{equation}
 \label{eXX}
  \omega_{p}^{(\pm)} = \pm \Omega + \sqrt{\left(\frac{I_{zz}\Omega}{2I_{rr}}\right)^2 + \omega_{0}},
\end{equation}
in which $I_{zz}$ and $I_{rr}$ are the moment of inertia around the rotating and radial
axes of the RW, respectively\cite{Izawa2008}. The two signs are for the forward and
backward branches. When $\Omega=66$~Hz (3.96~krpm) within the operation range, the
forward branch ($\omega_{p}^{(+)}$) matches $\sim$200~Hz. The study of the impact of
this mode is left for the spacecraft-level test (\S~\ref{s4}).

\section{Spacecraft-level tests}\label{s4}
\subsection{Measurements}\label{s4-1}
\begin{figure}[htbp]
 \centering
 \includegraphics[scale=0.5]{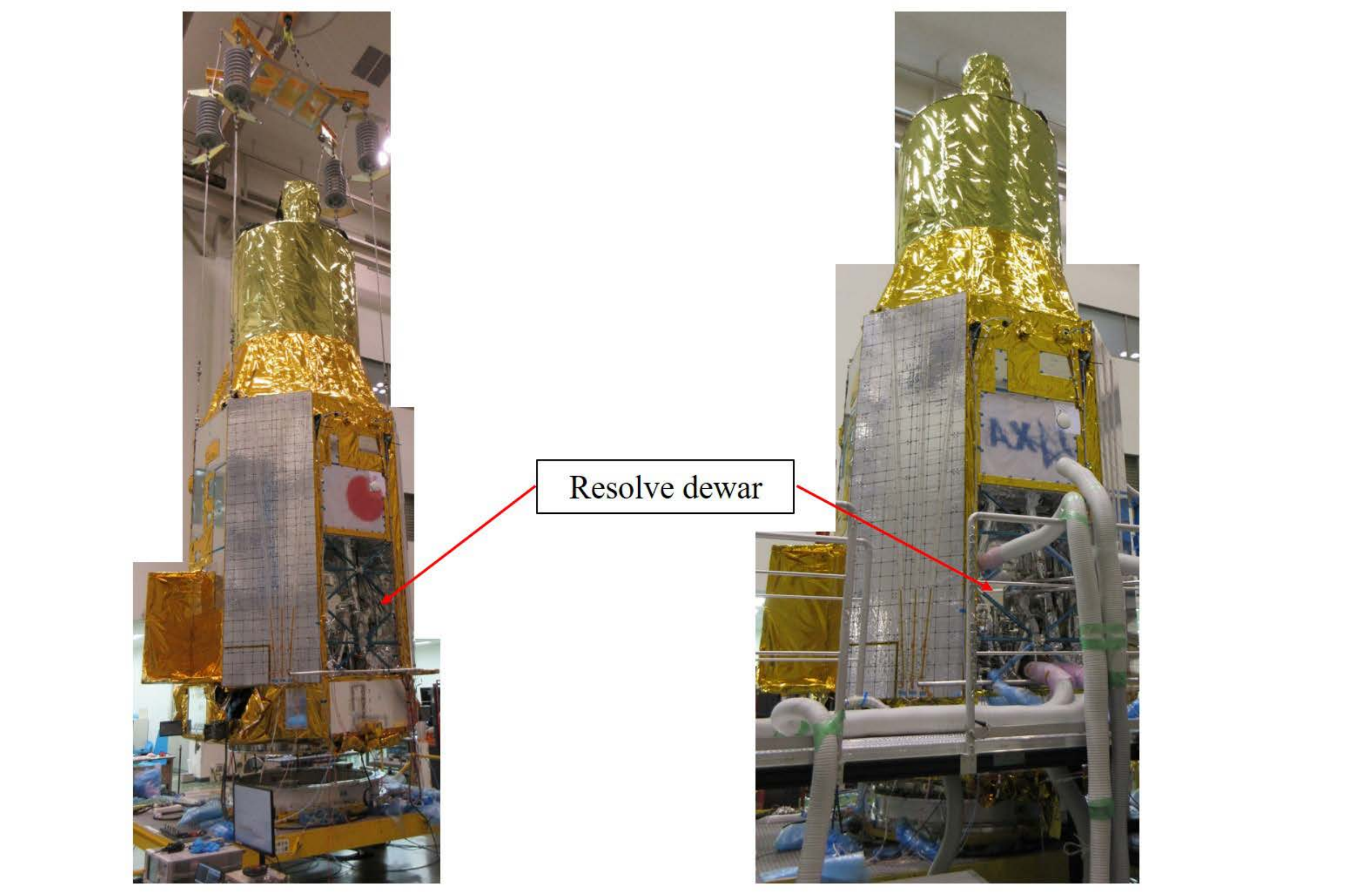}
 \caption{Spacecraft-level test configurations: (left) Spacecraft suspended and no
 \textit{Resolve} operation and (right) Spacecraft on the floor and \textit{Resolve}
 operation.}
 \label{fig:testpic}
\end{figure}

We performed the spacecraft-level tests twice in 2022 May 23--25 and June 10--11 in
different configurations. In the former, the entire spacecraft was suspended by a crane
for flight fidelity of the mechanical environment, but the \textit{Resolve} instrument
was not operated for the safety of the He plumbing (Figure~\ref{fig:testpic} left). In
the latter, the spacecraft was placed on the floor, but the detector was operated at
50~mK to make an end-to-end assessment (Figure~\ref{fig:testpic} right). The STC and JTC
were operated at 14.631 and 51.588~Hz after retuning the frequency prior to the
spacecraft integration. The two tests are hereafter called the suspended and unsuspended
tests.

In both tests, a total of 18 accelerometers (3 axes at 6 positions) were used to monitor
relevant positions of the spacecraft base plate, harness support bracket, SP1, and the
outermost shell of the dewar (Figure~\ref{fig:coordinate}). In the unsuspended test, the
detector noise spectra were additionally obtained on demand.

In the suspended test, we performed a full set of operations of the RWs and the
IRUs. Each RW was run up to the maximum 6.0~krpm and coasted down to 0, one by one,
sweeping over its entire frequency range. Each IRU was run at 155~Hz one by one, and a
combination of three out of the four for simulating the flight use. In the unsuspended
test, each RW was set at selected frequencies (1.5, 3.0, 3.96, and 6.0~krpm) to collect
the detector noise spectra. All RW units were coasted down from 6.0~krpm to 0. The start
was set apart by 30--60~min, so that their responses can be distinguished by the
concurrent RW rotating speeds. During the coast down, the detector noise spectra were
obtained on demand every 3 min to track the RW rotation changes within
the allocated data rate.

\subsection{Results}\label{s4-2}
\begin{figure}[!hbtp]
 \centering
 \includegraphics[width=0.9\columnwidth,clip]{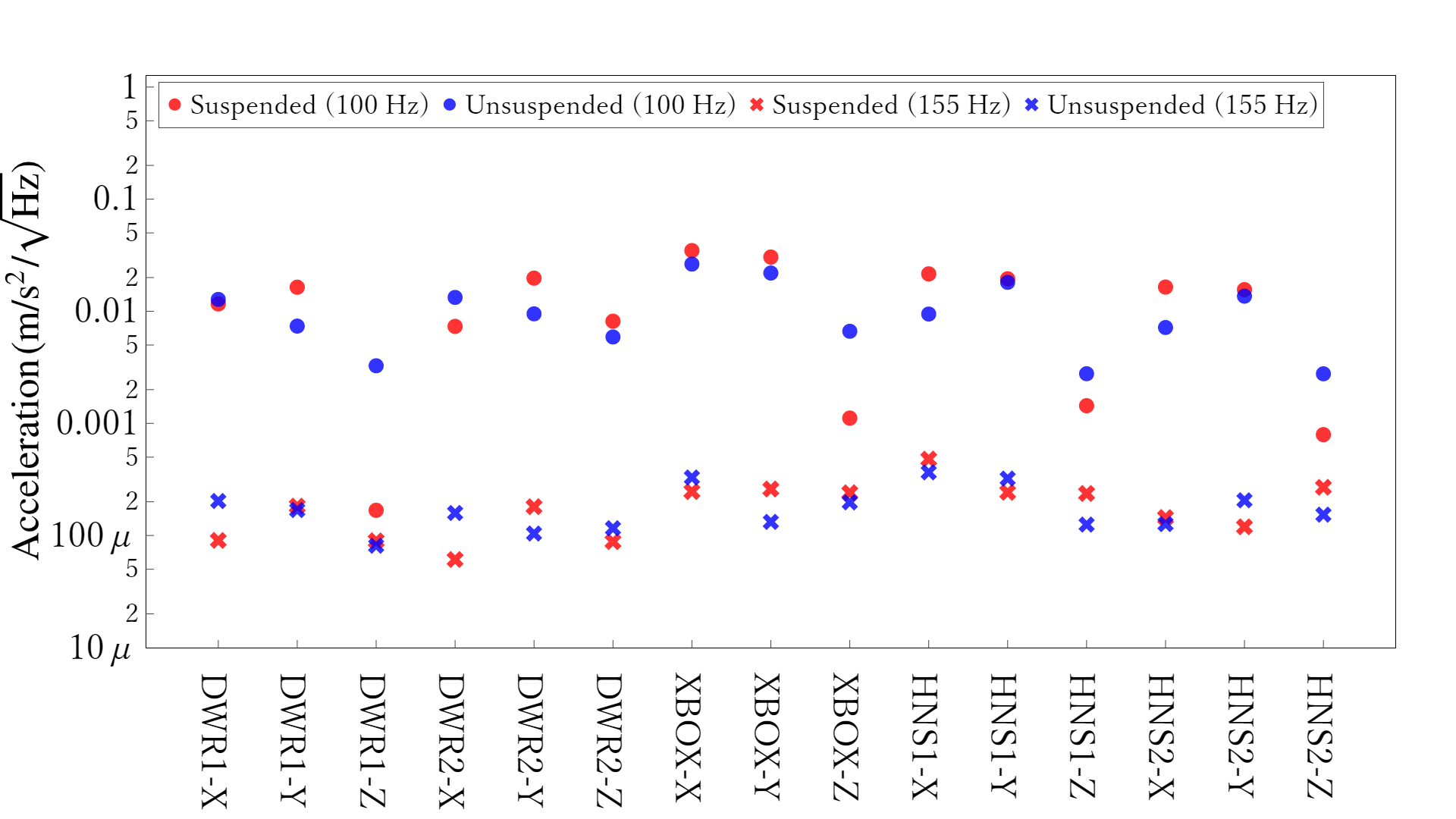}
 \caption{Comparison of accelerometer power between the suspended (red) and unsuspended (blue)
 tests for RW1 at 6.0~krpm and RW2--4 with no operation (circles) and an IRU operation (crosses).}
 \label{fig:comparison}
\end{figure}

First, we compare the accelerometer levels in the two configurations with and without
suspending the spacecraft. Figure~\ref{fig:comparison} shows the power in the
accelerometer spectra at 100 and 155~Hz respectively during the 6~krpm RW rotation and
the IRU operation. In the unsuspended test, the cryocoolers dominate the spectrum, but the
RW and the IRU signatures are distinctive at their fundamental frequencies. 
The power in the lines, from which the local continuum level was subtracted, varies from one accelerometer
channel to another, 
but there is no systematic
difference between the suspended and unsuspended
results. We can use the result of the unsuspended test for an end-to-end assessment with
sufficient flight fidelity.

\begin{figure}[htbp]
 \centering
 \includegraphics[width=0.9\textwidth]{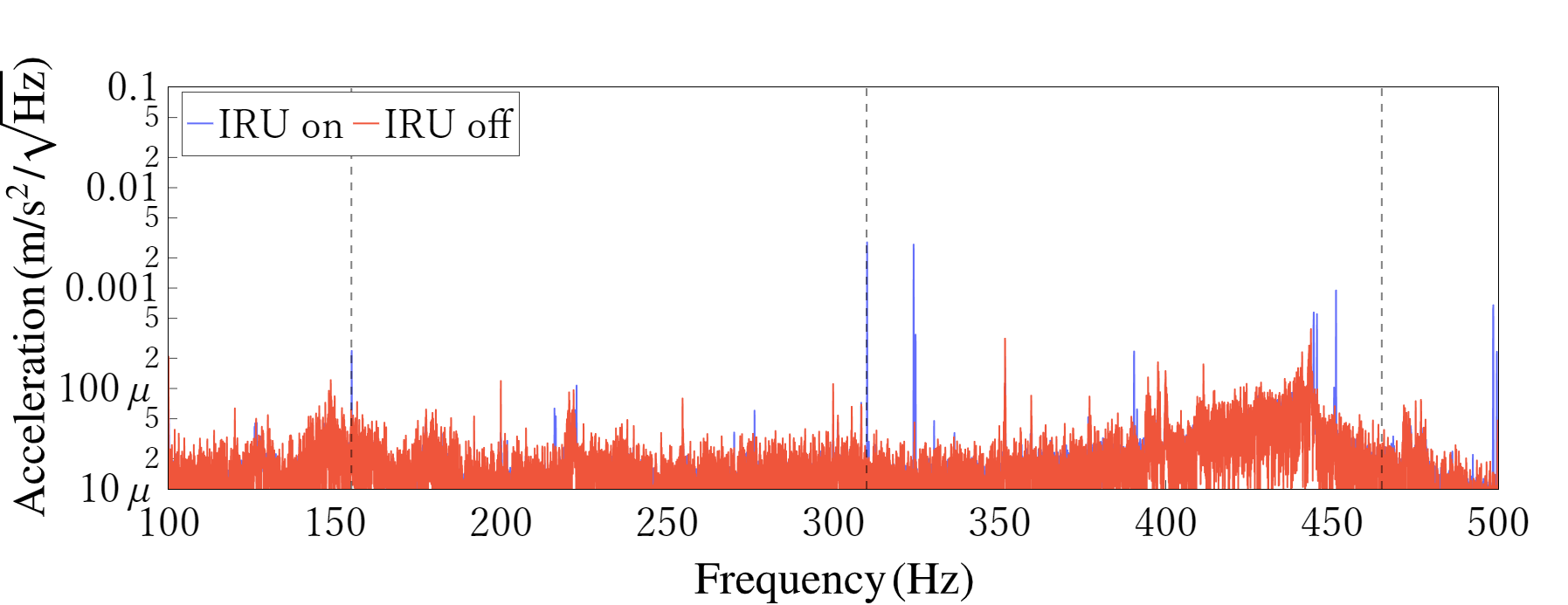}
 \caption{Accelerometer spectrum of IRU in comparison to the background during the
 suspended test for the HNS1-Z accelerometer. The fundamental frequency of 155 Hz and
 its harmonics are shown with dotted lines.}
 \label{fig:iru_acc}
\end{figure}

\begin{figure}[htbp]
 \begin{minipage}[b]{0.49\linewidth}
  \centering
  \includegraphics[height=0.4\textheight]{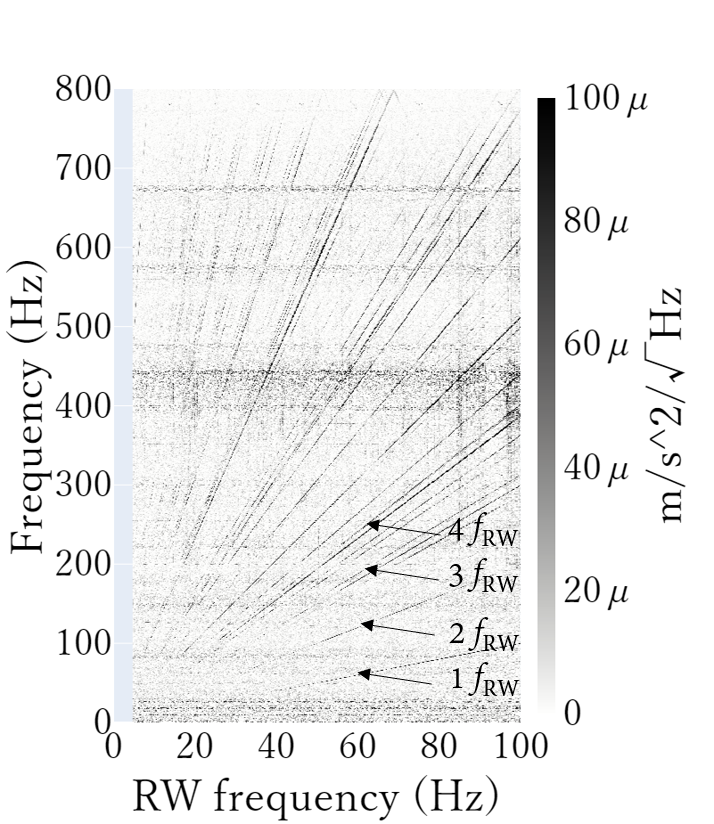}
  \subcaption{Accelerometer}\label{fig:rw_acc}
 \end{minipage}
 \begin{minipage}[b]{0.49\linewidth}
  \centering
  \includegraphics[height=0.4\textheight]{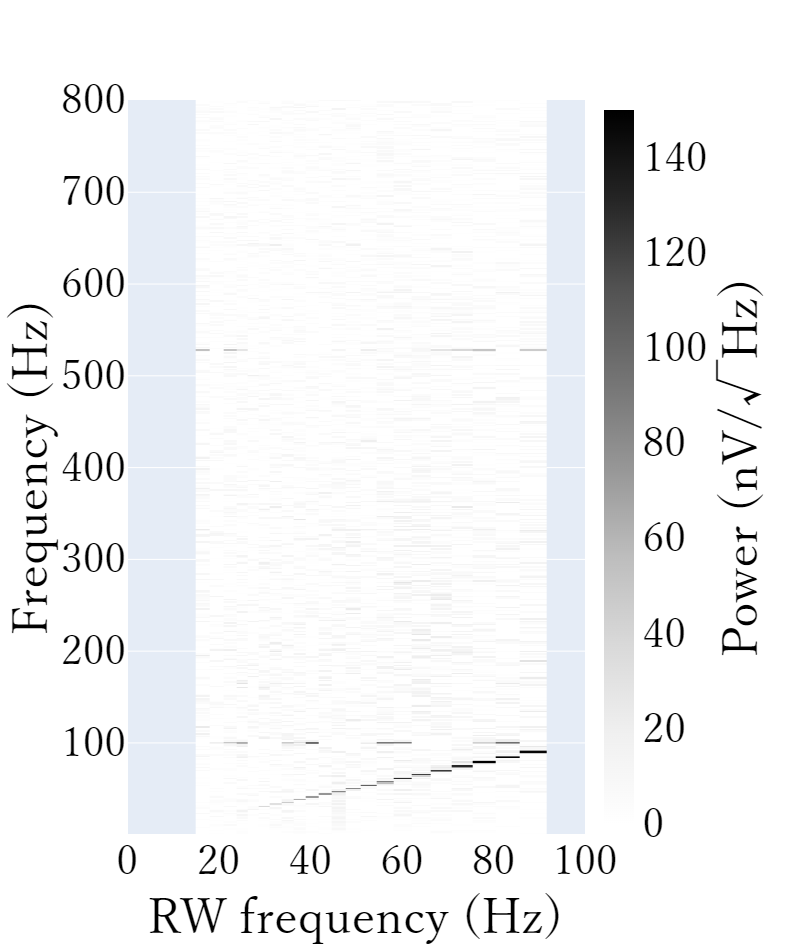}
  \subcaption{Detector noise}\label{fig:rw_nspec}
 \end{minipage}
 \caption{Spectrogram of the accelerometer spectra in the suspended test (left) and the
 detector noise spectra in the unsuspended test (right) during the RW1 coast down. The
 horizontal axis is translated from time to the concurrent RW rotation speed. The color
 scale is for the differential spectra by subtracting the median along the horizontal
 axis for the stationary component. Only positive signals are shown. The
 residual non-stationary horizontal line at 100~Hz is due to the RW2--4 that kept
 rotating at 6~krpm during RW1 coast down and that at 530~Hz is due to 10'th harmonic of
 JTC.}
\end{figure}

\begin{figure}[htbp]
 \centering
 \includegraphics[width=1.0\textwidth]{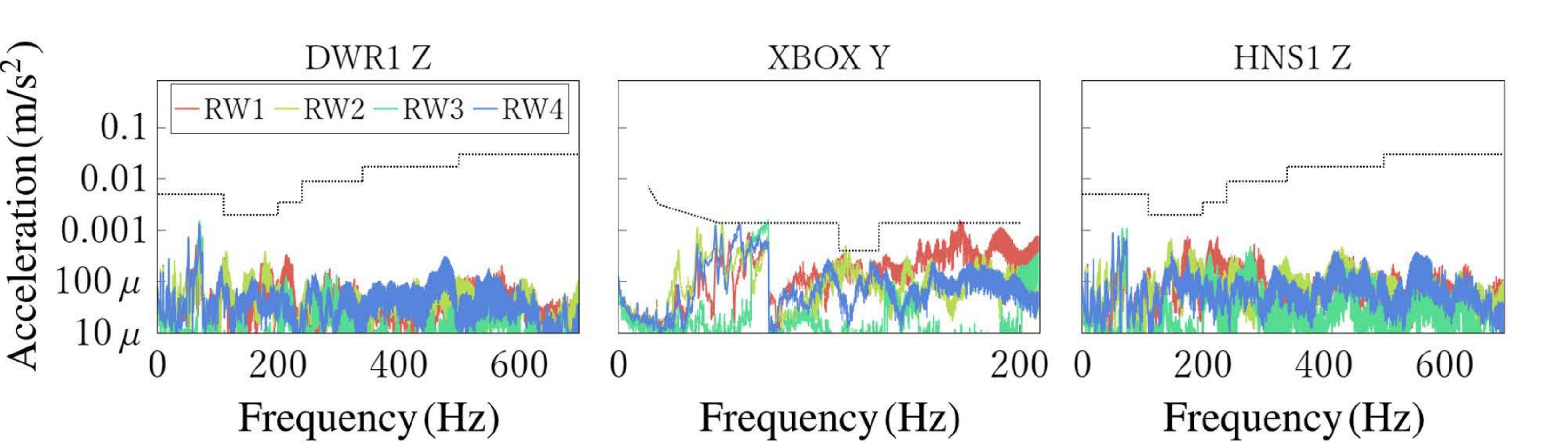}
 \caption{Strongest accelerometer spectra at each frequency measured during the RW
 coast down when the rotation is in its nominal operational range (1.5--4.5~krpm) for
 the dewer base plate (left), side panel 1 (center), and harness support bracket (right)
 in the suspended test for channels representing the interface axis. The dashed lines
 indicate the interface levels.}
 \label{fig:acc_lms}
\end{figure}

Next, we examine the accelerometer spectra of the IRUs and the RWs in the suspended
test, which is uncontaminated by the cryocooler lines, with a particular focus on the
non-diagonal components into high frequencies. Figure~\ref{fig:iru_acc} shows the
spectrum taken during the IRU operation compared to the background.  The fundamental
line at 155~Hz and its second harmonic (but not third and beyond) were observed.  The
acceleration at 155~Hz is below the interface level of the $1.4 \times
10^{-3}$~m~s$^{-2}$. Figure~\ref{fig:rw_acc} shows the accelerometer spectrogram as a
function of the RW rotation speed when the RW1 was coasted down. The line noise caused
by the RW should leave a trace that varies with time, or the concurrent RW rotation
speed, while the background line noise should leave a horizontal trace constant with
time. In fact, the constant background was removed in the plot by subtracting the median
of all spectra taken during the coast down. Besides the diagonal ($x=y$) traces by the
fundamental frequency, a dozen non-diagonal redistribution traces are recognized by the
higher order harmonics. There is no clear signature of the precession lines that would
leave curved traces in this representation. By taking the strongest signal at each
accelerometer frequency throughout the coast down, we can assess the worst level. The
result was below the interface level in the operating range (1.5--4.5~krpm) of all the
RWs (Figure~\ref{fig:acc_lms}).

\begin{figure}[htbp]
 \centering
 \includegraphics[width=0.9\textwidth]{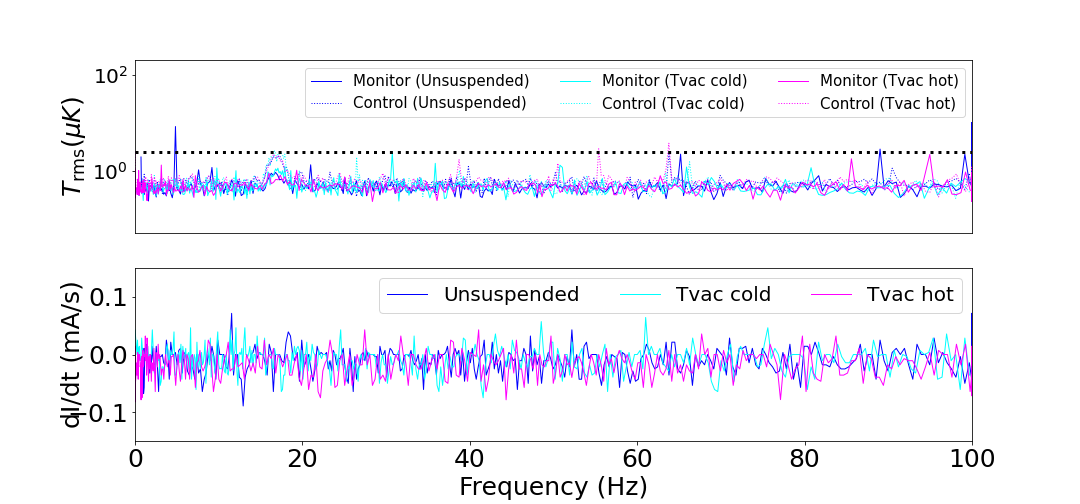}
 \caption{Same with Figure~\ref{fig:dwr_temp} during the RW1 coast down in the
 unsuspended test (blue), thermal-vacuum (Tvac) test cold and hot cases (cyan and
 magenta respectively). The spacing of the data is non-uniform along the RW, as the
 rotation decelerates faster at faster rotations during the coast down.}
 %
 \label{fig:rw_trms}
\end{figure}

We now explore the response of the \textit{Resolve} instrument using the data taken in
the unsuspended test. We examined the 50~mK stage temperature fluctuation
($T_{\mathrm{rms}}$) and the ADR magnet current consumption rate
($dI_{\mathrm{ADR}}/dt$) during the coast down. Figure~\ref{fig:rw_trms} shows the
result during the RW1 coast down. A positive response was found around 16~Hz, which was
unique to RW1. The RW1 response at $\sim$16~Hz was confirmed in the coast downs during
the spacecraft thermal-vacuum test held in 2022 August. No positive response was found
in the operating range of 25--75~Hz for all the RW units. We thus conclude that the
micro-vibration of the RWs does not cause a serious thermal load to the 50~mK stage
within its nominal operating range.

\begin{figure}[htbp]
 \centering
 \includegraphics[width=0.9\textwidth]{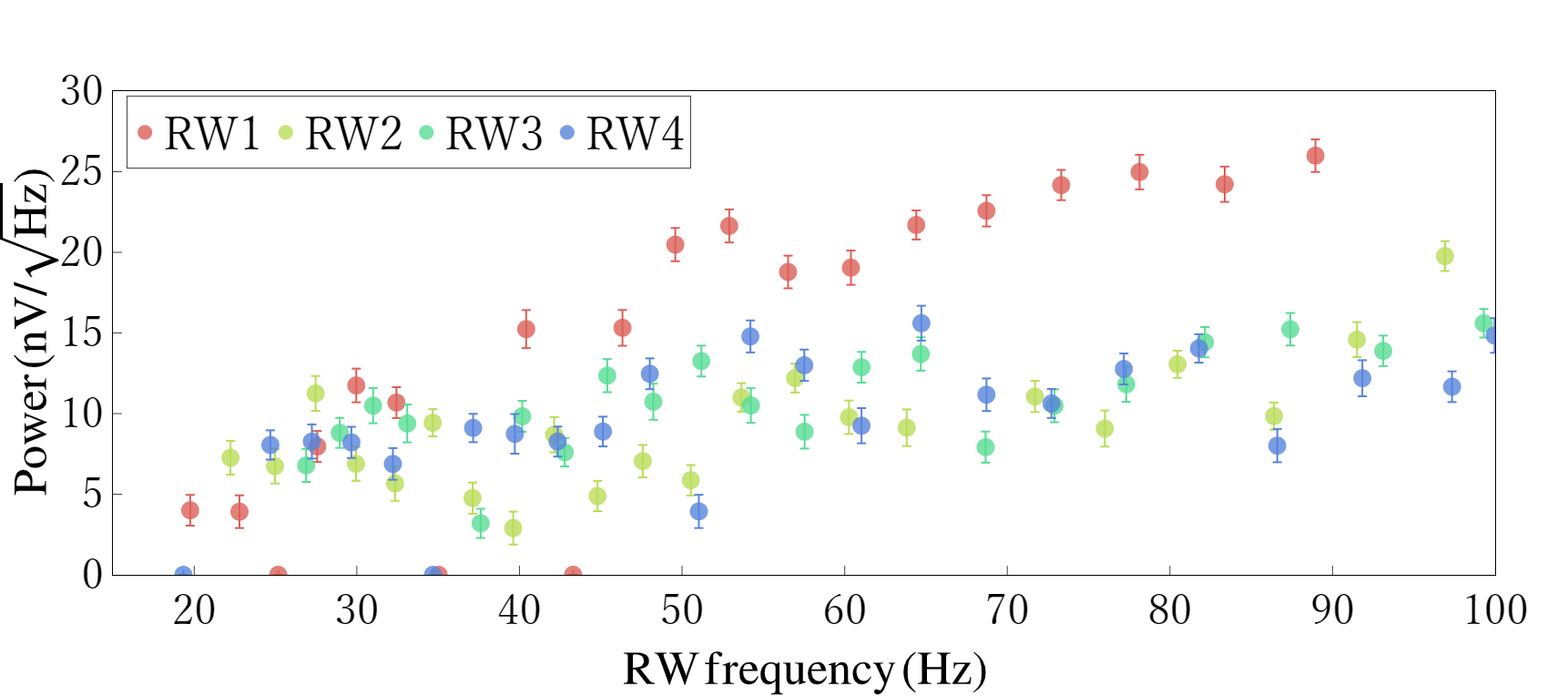}
 \caption{Line power of the detector noise spectra at the frequency of the concurrent
 rotating speed of each RW unit during coast down in the unsuspended test.}
 \label{fig:rw_nspec_trans}
\end{figure}

We next check the detector noise spectra taken during the coast
down. Figure~\ref{fig:rw_nspec} shows the spectrogram during the RW1 coast down using
all 36 microcalorimeter pixels for the sake of statistics. The constant component was
subtracted in the same way as Figure~\ref{fig:rw_acc}. The fundamental line of the RW
rotation speed is clearly seen with a decreasing intensity toward decreasing frequency,
but no obvious non-diagonal lines were found. We compared the power of the diagonal
component for each unit of RW1--4 during their coast down in
Figure~\ref{fig:rw_nspec_trans}. We found that the RW1 shows the largest response.

Finally, we make an end-to-end assessment in terms of the detector energy resolution. In
one of the RW coast down tests during the spacecraft thermal vacuum test, we illuminated
$^{55}$Fe X-ray sources over the entire detector array to take the X-ray spectra. Being
limited by the statistics, we compared the energy resolution during the former and
latter halves of the coast down and that before the coast down. We found no significant
changes in the Mn K$\alpha$ energy resolution among them. We also performed a long
($>$8~hr) $^{55}$Fe X-ray integration several times by operating the RW1--4 at a fixed
rate of 3~kprm and the magnetic torqer, which is a known electromagnetic interference
source\cite{Kurihara2022}, at some fixed duty ratio. 
Comparison with a reference measurement of no operation of these interference sources showed no significant difference in the performance of individual pixels. When the data of 33 pixels were combined, a barely significant difference of $<$0.02~eV for the Mn K$\alpha$ energy resolution of $\sim$4.0~eV.

\section{Summary}\label{s5}
We presented the result of 
a series of ground tests to verify the micro-vibration interference
from the spacecraft bus system to the microcalorimeter detector of \textit{Resolve}
onboard XRISM. In the instrument-level test, we injected micro-vibration as a function
of frequency with a force to yield the spacecraft-instrument interface level. We
evaluated the detector response using the 50 mK stage temperature stability and the ADR
magnet current consumption rate. We found the strong responses at $\sim$ 200, 380, and
610 Hz. In the former two cases, the beat among the injected frequency and the
cryocooler frequency harmonics of the JTC and the STC are observed in the detector noise
spectra. In the spacecraft-level test, we operated the two major sources of the
micro-vibration in the spacecraft ---the RWs and the IRUs--- with suspending the entire
spacecraft for the mechanical flight fidelity and without suspending the spacecraft and
with operating \textit{Resolve} for an end-to-end assessment. We observed numerous noise
lines besides the fundamental frequencies of the RWs and IRUs, but none of them cause an
unacceptable level of response in the instrument within the nominal operational ranges
of the RWs and IRUs. We conclude that the verification of the micro-vibration is
complete and the risk is sufficiently mitigated for the flight.

It would have been difficult to assess the micro-vibration interference outcome before we
actually did an end-to-end test, especially when we have non-diagonal injection and
response as shown here. In case of XRISM, it was fortunate that none of the
micro-vibration sources in the spacecraft
brought noise at frequencies that the
detector is sensitive to. 
If it were the case, we had an optional solution to change the
nominal frequency range of RWs. Such flexibility would be worth considering in the
design of future missions.

\acknowledgments
This work is made possible only with  significant contributions from all the XRISM
\textit{Resolve} team members and the engineers of SHI and NEC.

\bibliography{main} 
\bibliographystyle{spiejour} 

\vspace{2ex}\noindent\textbf{Takashi Hasebe} is a post doctoral fellow of the Kavli
Institute for the Physics and Mathematics, the University of Tokyo. He received his Ph.D.
degree in physics from Hiroshima University in 2016. His current research interests
include cryogenics and millimeter-wave optics.

\listoffigures

\end{spacing}
\end{document}